\documentclass[12pt,preprint]{aastex}
\usepackage{natbib}
\usepackage{graphicx}
\begin{document}

\title{A Catalog of \ion{H}{1} Clouds in the Large Magellanic Cloud}
\author{S. Kim\altaffilmark{1}, E. Rosolowsky\altaffilmark{2}, Y. Lee\altaffilmark{3},
Y. Kim\altaffilmark{1}, Y.C. Jung\altaffilmark{4}, M.A. Dopita\altaffilmark{5},
B.G. Elmegreen\altaffilmark{6}, K.C. Freeman\altaffilmark{5}, R.J. Sault\altaffilmark{7},
M. Kesteven\altaffilmark{7}, D. McConnell\altaffilmark{7}, \& Y.-H. Chu\altaffilmark{8}}
\altaffiltext{1}{Department of Astronomy \& Space Science, Sejong University,
KwangJin-gu, KunJa-dong 98, Seoul, 143-747, Korea; e-mail: sek@sejong.ac.kr}
\altaffiltext{2}{Harvard-Smithsonian Center for Astrophysics, 60
Garden St., MS-66, Cambridge, MA 02138, USA}
\altaffiltext{3}{Korea Astronomy and Space Science Institute,
Daejeon, 305-348, Korea} \altaffiltext{4}{Physics Department, Louisiana State University,
LA 70803, USA} \altaffiltext{5}{Mount Stromlo Observatory, Weston
Creek, PO Box, ACT 2611, Australia} \altaffiltext{6}{IBM Research Division, T.J. Watson Research Center, P.O. Box 218, Yorktown Heights, NY 10598, USA} \altaffiltext{7}{Australia
Telescope National Facility, Epping 76, NSW2121, Australia}
\altaffiltext{8}{Astronomy Department, University of Illinois at Urbana-Champaign,
West Green St., IL61801, USA}

\begin{abstract}
A 21 cm neutral hydrogen interferometric survey of the Large
Magellanic Cloud (LMC) combined with the Parkes multi-beam
\ion{H}{1} single-dish survey clearly shows that the \ion{H}{1} gas
is distributed in the form of clumps or clouds.  The \ion{H}{1}
clouds and clumps have been identified using a thresholding method
with three separate brightness temperature thresholds ($T_b$). Each
catalog of \ion{H}{1} cloud candidates shows a power law
relationship between the sizes and the velocity dispersions of the
clouds roughly following the Larson Law scaling $\sigma_v \propto
R^{0.5}$, with steeper indices associated with dynamically hot
regions.  The clouds in each catalog have roughly constant virial
parameters as a function mass suggesting that the clouds are all in
roughly the same dynamical state, but the values of the virial
parameter are significantly larger than unity showing that turbulent
motions dominate gravity in these clouds.  The mass distribution of
the clouds is a power law with differential indices between $-1.6$
and $-2.0$ for the three catalogs. In contrast, the distribution of
mean surface densities is a log-normal distribution.
\end{abstract}

\section{Introduction}

It is well known that interstellar turbulence is a vital element of
theories of star formation. During the last decade, theoretical models
of the importance of turbulence on star formation have been produced
(Scalo and Elmegreen 2004, for a review) using observations to provide
constraints to the models. MacLow and Klessen (2004) stress that the
size line width scaling relation can arise from the fact that
molecular clouds are hierarchical structures in turbulent flow and
theory of turbulence leads to a log-normal probability gas density
function (Padoan and Nordlund 2002).

%REVISE
A 21 cm neutral hydrogen interferometric survey of the Large
Magellanic Cloud (LMC) (Kim et al. 2003) combined with the Parkes
multi-beam \ion{H}{1} single-dish survey (Staveley-Smith et al. 2003)
clearly shows that the \ion{H}{1} gas is distributed in the form of
clumps or clouds as well as shells (Kim et al. 1999). These features
are also well demonstrated in the ATCA survey alone (Kim et al. 1998).
Wada et al. (2000) present that high-resolution 2D numerical
simulations of the interstellar atomic hydrogen gas in the LMC and find
that the statistical properties of \ion{H}{1} from their simulation are
similar to those of observed \ion{H}{1} from the ATCA interferometry
survey.

In this present study, we present catalogs of various clump-like or
cloud-like features in the LMC using the combined ATCA$+$Parkes
survey.  The LMC is the nearest disk galaxy to the Milky Way but its
face-on geometry avoids the complications of the line-of-sight
blending that pose difficulties in studying discrete \ion{H}{1} clouds
in the Milky Way.   This paper presents the properties of the
\ion{H}{1} clouds and compares those properties to the scaling
relationships for the atomic and molecular gas in the Milky Way.

\section{The Cloud Catalog}

This section describes the selection criteria for the \ion{H}{1} cloud
candidates chosen from the \ion{H}{1} data cube of the LMC. The \ion{H}{1}
clouds were identified using a brightness temperature threshold ($T_b$)
applied to the 21-cm neutral hydrogen gas survey of the LMC.  We
cataloged the data from the combined surveys (Kim et al. 2003) of the
Australia Telescope Compact Array (ATCA) (Kim et al. 1998) and the Parkes
64-m single-dish telescope (Staveley-Smith et al. 2003).
%REVISE
The results of \ion{H}{1} aperture synthesis mosaic survey were made by combining
data from 1344 separate pointing centers and the Parkes 64-m multi-beam data
correspond to 11$^{\circ}$.1 $\times$ 12$^{\circ}$.4 on the sky. The spatial
resolution of the map is 50$''$ $\sim$ 55$''$ corresponding to a physical
resolution of 12 $\sim$ 14 pc at the distance of the LMC (Alves 2004).
The observing band was centered on 1.419 GHz, corresponding to a central
heliocentric velocity coverage from $-$33 to $+$627~km~s$^{-1}$ with a velocity
resolution of $\delta v \equiv 1.649$~km~s$^{-1}$. Using the technique, a variant
of approach described by Schwarz and Wakker (1991), a composite image was formed
by filtering out the short spacing data from the ATCA image and then adding the
Parkes image. The Fourier transformed Parkes images were added to the final images
with no weighting in {\sc MIRIAD} task {\sc IMMERGE}. The RMS noise level of the
final cube is $\sigma_{rms}$=2.5~K.

We identify clouds using a brightness temperature threshold and
identifying all connected pixels above the temperature threshold as
belonging to a single cloud \citep[cf.][]{dect86,sol87,syscw}.
Ideally, one would like to define clouds with a zero threshold
intensity. However, low threshold intensities are impractical in view
of the noise level in the spectra and more importantly because of the
blending of adjacent clouds which often occur in crowded regions. On
the other hand, with too high a threshold intensity, regions are
severely truncated, and it is impossible to obtain a reliable estimate
of the sizes and velocity dispersions, and related parameters
(Scoville et al. 1987).
%REVISE
Identification of the clouds are conducted within {\sc IRAF}\footnote{Image
Reduction and Analysis Facility is written and supported by the IRAF programming
group at the National Optical Astronomy Observatories.} with the computer
algorithm we have written in {\sc FORTRAN}. In order to make the FORTRAN code
available in the environment of {\sc IRAF}, a Command Language (CL) for {\sc IRAF}
is used to incorporate compiled code into the CL by calling newly added functions
within a CL program.
%Identification of the clouds are conducted within with the computer algorithm
%written in IRAF. To make a FORTRAN program using
%IMFORT into a IRAF task, a CL foreign task interface is required to
%connect the program to CL callable task.
We arbitrarily simplify the CL file for easy handling. One merit of making a task
within {\sc IRAF} is that every data can be handled in {\sc IRAF} image form, which
can be transformed to FITS form easily, and can be transported to other reduction
packages, if necessary.  Moreover, the identified clouds can be separately
regenerated for further analysis within {\sc IRAF}. When making a catalog of
identified clouds, one may adjust the threshold pixel numbers, as one may not want
to include the minimum number of pixels for defining a cloud.  Along with
identifying clouds, the algorithm calculates intensity-weighted mean values of
longitude, latitude, and velocity as well as these the dispersions of these
parameters.  In addition, the total integrated intensity, the number of identified
clouds, and the number of pixels involved are also reported.

We adopt three separate temperature thresholds: $T_{thresh}$=64, 32,
and 16~K, identifying 195, 406, and 468 clouds at each threshold
level. A pixel identified as part of a cloud at high thresholds will
also be included in catalogs at lower thresholds (cf. Scoville et
al.~1987).  The catalogs respectively contain 3, 19 and 37 \% of the
total luminosity in the data cube. We do not apply a minimum size
criterion to the clouds included in the catalog, so objects as small
as 1 pixel are nominally included though few objects this small are
identified (see \S \ref{mspec-section}).

For a cloud consisting of a set of pixels with positions
$\{x_i,y_i,v_i\}$ and intensity $\{I_i\}$.  The mean velocity for a
cloud is calculated as the intensity weighted mean over all the pixels
within a cloud:
\begin{equation}
\langle v\rangle = \frac{\sum_{i}\, v_{i} I_{i}}{\sum_{i}\, I_{i}}
\end{equation}
where $I$ is intensity of a pixel and $i$ runs over all pixels in the
cloud. The velocity dispersion ($\sigma_{v}$) is also
intensity-weighted:
\begin{equation}
\sigma_{v}^{2} = \frac{\sum_{i}\, I_i (v_{i} - \langle v \rangle)^{2}}
{\Sigma I_i}.\label{veldisp}
\end{equation}
where $\Sigma I_i$ is the integrated intensity of pixels within the
cloud.

We have also observed the internal velocity dispersions and sizes of
the identified \ion{H}{1} clouds. We define the internal velocity dispersion as
the rms magnitudes of gas motions along individual lines of sight (at position $x,y$)
with respect to the average cloud velocities along those directions. The observed
dispersion of each line of sight, $\sigma(x,y)$, is derived from:
\begin{equation}
\sigma^2(x,y)=\frac{\Sigma_k I_k[v_k - v_c(x,y)]^2}{\Sigma_k I_k}
\label{eqn3.1}
\end{equation}
where $v_c(x,y)$ is the centroid velocity at position ($x,y$) and the
sum $k$ runs over all pixels in the cloud with position ($x,y$).  The
centroid velocity at this position is:
\begin{equation}
v_c(x,y)  = \frac{\Sigma_k I_k v_k}{\Sigma_k I_k}. \label{eqn3.2}
\end{equation}
The internal velocity dispersion of the entire cloud is then given by
\begin{equation}
\sigma^2 = \frac{\Sigma_{ij}[\sigma^2(x_i,y_j)] W_{ij}}{\Sigma_{ij} W_{ij}}.
\end{equation}
where the summation $\Sigma_{ij}$ runs over all $x,y$ positions in the
data cube and $W_{ij}$ is the integrated intensity for each position.
The internal velocity dispersion is an attempt to remove the effects
of the point-to-point velocity differences in the centroid of the
cloud (such as velocity gradients in the cloud). Therefore, the total
velocity dispersion (Equation \ref{veldisp}) is a mixture of the
large-scale velocity variations across the region and the smaller-scale
motions responsible for the line width, since the thermal contribution is
negligible.

% ER REVISED.
The size of each cloud is calculated from based on its extent in the
spatial dimension.  We measure the second moment of the emission
around the centroid of emission:
\begin{equation}
\sigma_{x}^{2} = \frac{\sum_{i}\, I_i (x_{i} - \langle x \rangle)^{2}}
{\Sigma I_i}.\label{posdisp}
\end{equation}
with a similar expression for $\sigma_y$.  The radius (size) of a cloud is
then reported as
\begin{equation}
R = \left(2\ln 2~\sigma_x \sigma_y \right)^{1/2}
\end{equation}
We make no correction to the size for beam convolution effects.  Since
the radius is calculated from a dispersion, clouds must be at least
two pixels across in both directions.

The mass of the cloud is determined by its luminosity in the 21-cm line with
no corrections for optical depth effects.
%REVISE
For an integrated intensity $W$ measured in K~km~s$^{-1}$,
\begin{equation}
M_{cloud} = 1.822\times 10^{18} W A_{cl} m_{\mathrm{H}}
\end{equation}
where $A_{cl}$ is the total area of the cloud in cm$^2$. No correction
for the presence of helium or metals is included in the mass measurement.
%REVISE
The potential uncertainties in estimating masses of the \ion{H}{1} gas
clouds include the importance of \ion{H}{1} self-absorption correction
to the \ion{H}{1} column densities and the existence of molecular gas
clumps not traced by CO within the apparent diffuse \ion{H}{1} gas
(Grenier et al. 2005).  Given that $N_H/\tau \approx T_s ~\Delta V$,
where $N_H$ is column density, $T_s$ is spin temperature, $T_s$ and
$\tau$ cannot be independently determined (Dickey and Lockman 1990).
Thus, it is not easy to determine a definite $N_H$ in the optically
thick case from the line analysis.
%% However, simplifications can be introduced in
%% order to correct the detected \ion{H}{1} line flux using the isophote
%% axial ratio as the measure of disk inclination (Giovanelli et
%% al. 1994). The correction factor, $(a/b)^{0.12}$, for the disk of the
%% LMC is close to 1, and this effect is small with less than 10\% of the
%% \ion{H}{1} line flux self-absorbed.

The physical parameters of the individual \ion{H}{1} cloud candidates
are reported for each catalog in Tables 1 through 3. Between the
catalog, the \ion{H}{1} clouds are similar in structure, but appear to
be inhomogeneous and clumpy in nature as seen in Figure 1 and Figure 2.
\ion{H}{1} clouds often express similar pattern to those of molecular clouds
(Heiles 1997).  Large clumps within molecular clouds have been observed to
contain a number of smaller clumps moving randomly (Myers 1978; Blitz 1980).
For decades, observations of molecular clouds have shown that the internal
velocity dispersion of each region is well correlated with its size
and mass; this is known as the Larson's scaling law (Larson 1981). Observations
of non-thermal line widths indicate the turbulent nature of the interstellar
clouds (Falgarone and Phillips 1990).
%% XXX What's the point of the above paragraph?

%REVISE

\section{Discussion}
\subsection{Distribution of \ion{H}{1} cloud candidates}

This study is directed at revealing the cloudy nature of the atomic
interstellar medium (ISM) in the LMC. The locations of identified
\ion{H}{1} cloud candidates are shown in Figure 1. In Figure 1, we
plot the location and sizes of \ion{H}{1} clouds in the LMC. The LMC
is nearly face-on ($i=22^{\circ}-33^{\circ}$) and the synthesized beam
is 50$''-$55$''$ in diameter, which corresponds to a spatial
resolution of 12$-$14 pc in the disk of the LMC at the distance of the
LMC (50.1 kpc, Feast 1991; Alves 2004). It is interesting to note
smaller clouds identified with a higher brightness temperature
threshold inside a big cloud with a lower brightness temperature
threshold. This feature might indicate the hierarchical structure of
the neutral hydrogen gas in the ISM.
%Since the LMC is nearly face-on with the
%inclination angle of 22$^{\circ}$-33$^{\circ}$ and the synthesized
%beam of present \ion{H}{1} survey is 50-55$''$ in diameter, the
%smallest \ion{H}{1} cloud we can resolve corresponds to a size of
%12$-$14 pc at the distance of the LMC (Feast 1991).
We note that the sizes of the \ion{H}{1} cloud candidates do not depend on their
locations in general. Regardless of the brightness temperature threshold, a total of
87 \ion{H}{1} cloud candidates lie close to the northern \ion{H}{1} spiral arm and the
other 159 \ion{H}{1} cloud candidates lie close to the southern \ion{H}{1} spiral arm.
On the other hand, 823 \ion{H}{1} cloud candidates are located in the other regions of
the LMC.
%% XXX Each cloud entry in the catalog should be tagged with what
%% region it belongs to.  Or the regions should be indicated in a
%% figure.  Or both.
In contrast to the $^{12}$CO emitting clouds in the $^{12}$CO $J=1\rightarrow0$
map of the LMC by the NANTEN 4-m telescope (Fukui et al. 2001), the \ion{H}{1}
cloud candidates identified here in the present study distribute quite uniformly.
In general, there is no major concentration of \ion{H}{1} cloud candidates in
any part of the LMC (Figure 2).  We highlight several regions of the
LMC in Figure 2 associated with dynamically ``hot'' regions like 30
Doradus or the supergiant shell LMC 4 (Meaburn 1980) and label
associated clouds in Tables 1--3.

%REVISE
%COMPARE THE LARSON LAW AND THE SIMULATION RESULTS!!
For a given temperature threshold, \ion{H}{1} cloud candidates
both in the dynamically hot regions and in the other regions obey
the power law relationship between the size $R$ and the velocity
dispersion $\sigma$ of \ion{H}{1} cloud candidates as shown in
Figures 4, 5, and 6. These phenomena are similar to the Larson Law
in its characterization of molecular clouds. The dynamically hot
regions are characterized by quantitative measurements of
turbulence. The velocity difference map (Figure 8 in Kim et al.
1998) generated from the modulus of the velocity difference
between the peak radial velocity and the velocity smoothed over to
an effective resolution of 20$'$ can help us to quantify dynamically
hot effects in the ISM. The velocity dispersions $\sigma$ of the \ion{H}{1}
clouds increase with their sizes $L$ as power law, $\sigma \propto L^{\alpha}$.
A slope $\alpha$ of velocity dispersion to size relation of \ion{H}{1} cloud
candidate in the dynamically hot regions varies logarithmically from 0.33$\pm$0.14
to 0.88$\pm$0.32. A slope of the power law relationship between velocity dispersion
and size appears to be similar as 0.46($\pm$0.06)$-$0.52($\pm$0.04) in the relatively
quiet regions. Figure 7 presents a slope of the power law between the size and the
velocity dispersion in the entire regions of the LMC for a given brightness
temperature threshold by which the \ion{H}{1} cloud candidates were identified and
selected. It is suggested that the slope of size$-$velocity dispersion relation
varies from 0.48 to 0.53, depending on the locations of clouds and the temperature
threshold. In regions other than 30 Doradus, the spiral arms, LMC 4, and the western
end of bar, the slope does not vary much with the temperature threshold. The
arithmetic mean value of slopes seen in Figure 7 is about 0.5$\pm$0.02 which seems
to match values obtained by other studies in the statistical uncertainties and
similar to the Kolmogorov value. Our derived slope in some of dynamically hot
regions for a given brightness temperature threshold is significantly higher than
exponent of Larson size-line width relation, 0.38.
%For example, our derived slope in some of dynamically hot regions for a given
%brightness
%temperature threshold is slightly higher than the slope value of 0.38 first identified
%by Larson (1981) and similar to the Kolmogorov value.

In recent studies (Hennebelle et al. 2006; V\'azquez-Semadeni et al. 2006), synthetic
\ion{H}{1} spectra have been computed and the CNM clouds are formed in their numerical
simulations. In their simulations, the CNM clouds are formed by dynamical compressions
in the WNM, appear to have velocity dispersons of about 1 km~s$^{-1}$, and follow
Larson-type relations. They find that the velocity dispersion increases with the size of
the CNM structures, $L$, as $\sigma \propto L^{0.4}$. Their velocity dispersions are
comparable to the values observed in \ion{H}{1} clouds by Heiles \& Troland (2003). In
our case, cold and warm gas are mixed. As expected, typical velocity dispersions are
1.5$-$4 times larger than theirs depending on the threshold used to define the clouds.

\subsection{The Mass Spectrum of \ion{H}{1} Cloud Candidates}
\label{mspec-section}
In Figure \ref{mspec}, we plot the cumulative mass distribution of
cloud candidates for the three catalogs given in Tables 1 to 3.
The cumulative mass distribution plots the number of clouds with a
mass greater than a given mass.  The cumulative mass distribution is
the integral of common differential mass distribution ($dN/dM$):
\begin{equation}
N(M'>M) = \int_M^\infty \frac{dN}{dM'} dM'
\end{equation}
We choose the cumulative mass distribution rather than the
differential distribution because (1) estimates of the differential
mass distribution are prone to systematic errors due to binning choice
and (2) fits to the cumulative mass distribution can account for
uncertainties in the cloud mass which analyses of the differential
mass spectrum commonly neglect (Rosolowsky 2005).  For each catalog,
we fit a power law function  to the cumulative mass distribution.
\begin{equation}
N(M'>M) = N_0\left(\frac{M}{10^5 M_{\odot}}\right)^{\gamma+1}
\label{mspeceqn}
\end{equation}
$N_0$ is the number of clouds in the derived distribution with masses
larger than $10^5~M_{\odot}$ and $\gamma$ is the index of the
differential mass distribution.  For $\gamma>-2$, the mass
distribution is top heavy and most of the mass is found in the high
mass clouds.  For $\gamma<-2$, the opposite is true; and $\gamma=2$
implies and equal amount of mass in every logarithmic bin of the
mass distribution.

% ER REVISED
For each mass distribution, we estimate the parameters of the mass
distribution using the algorithm of Rosolowsky (2005).  The adopted
algorithm is a non-linear regression of the data to Equation
\ref{mspeceqn} using the error-in-variables method \citep{errinvar}
for parameter estimation.  The uncertainties in the cumulative mass
function are given by counting errors and the uncertainties in the
mass function are assumed to be 5\%.  We report the derived parameters
for the mass distributions of the three catalogs in Table 4. For each
catalog, we adopted a different completeness limit appropriate for
that catalog and reported this value in Table 4. Since clouds down to
the size of 1 spatial resolution element by 1 velocity channel are
included in the catalog, it would appear that the completeness limit
is set by our brightness threshold:
\begin{equation}
M_{comp} = 0.014~\ell^2~\delta v~T_{thresh} M_{\odot}
\end{equation}
where $\ell$ is the linear extent of 1 pixel in pc and $\delta v$ is
the channel width in km~s$^{-1}$.  For a pixel size of 9.7 pc and a
channel width of 1.649 km~s$^{-1}$, the completeness limit is 3.7, 7.3
and 14.6~$M_\odot$ for $T_{thresh}=16,32,$ and 64 K
respectively. Inspection of Tables 1 to 3 or Figure \ref{mspec} shows
no clouds with such a low mass are found in the catalog, though they
should be detectable.  Such clouds are not detected because the ISM of
the LMC does not contain clouds with these properties.  We find that
the clouds in the LMC are all drawn from populations with roughly
constant virial parameter (see below) and clouds with such a low mass
would have sizes and line widths significantly smaller than the beam
width and channel size and thus be undetectable given the data.

%ER REVISED
The adopted limits are, instead, inferred from extrapolations of the
dynamical states (virial masses) of the clouds down to a dynamical
completeness limit established by the resolution of the observations.
The dynamical mass estimate for a cloud is given by
\begin{equation}
M_{VT} = \frac{5 R\sigma_v^2}{\alpha G}
\end{equation}
where $\alpha$ is the virial parameter for the clouds.  If the clouds
are self-gravitating and complicating effects such as magnetic fields
or external pressure are negligible, $\alpha < 2$.  Virialized clouds
have $\alpha \approx 1$.  In Figure \ref{vircomp}, we plot the
dynamical mass estimate $(5R\sigma_v^2/G$, i.e.~the virial mass with
$\alpha$ assumed to be unity) as a function of the total luminous mass
($\mu M_{21cm}$, where $\mu=1.4$ is the mean particle mass accounting
for metals and helium).  The cloud populations stretch out along loci
of constant virial parameter $\alpha$.  We plot the median virial
parameter for each catalog as a gray line.  For a 64, 32 and 16~K
threshold, we find median virial parameters of $\langle \alpha \rangle
= 7.2, 13.9,\mbox{ and } 28.2$ respectively.  The apparent
differences in virial parameters for the cloud catalogs are a
consequence of changing the threshold contour $\langle \alpha \rangle
\propto T_{thresh}^{-1}$ with no attempt to account for emission below
the threshold level.  Since the mean virial parameters are
significantly larger than unity, self-gravity is negligible for these
clouds and their internal energetics are dominated by their kinetic
energy (See \S \ref{virparam}).

% ER REVISED
The distribution of dynamical mass estimates is truncated at the
dynamical completeness limit given by the horizontal dotted line.
This limit is defined as the dynamical mass limit for $R = 2\ell =
19.4$~pc and $\sigma_v = 0.5 R^{0.5}$ = 2.20~km~s$^{-1}$.  Note, we
have used the velocity dispersion implied by size-line width
relationship (Figure \ref{rdv}) since is this larger than what would
be implied by the channel width [$\sigma_v = \delta v / (8 \ln
2)^{1/2} =$~0.71~km~s$^{-1}$]. Thus, it is the spatial resolution
which defines how deep into the LMC cloud population the catalog
samples.  Similarly, the luminous mass distribution is truncated at
the vertical dotted line established by the threshold temperature and
the minimum radius and line width given above.  We adopt the
completeness limits given in Table \ref{mspec-index} so that the
completeness limit is above where the lower envelope of the data
intersects with the dynamical completeness limit and the left-hand
envelope of points is similarly above the intersection with the
luminous mass completeness limit.  Points found below the dynamical
completeness limit have unreliable estimates of their sizes or line
widths because these cloud properties are affected by pixelization.

Returning to the mass distributions, we find that a power law
distribution is a good model for all but the most massive clouds in
each catalog.  For the higher thresholds (64 and 32 K), the index of
the mass distributions is $\gamma \approx -1.7$, but the distribution
significantly steepens to be consistent with $\gamma \approx -2.0$ at
lower thresholds.  Unlike the mass distributions of GMCs in the LMC
(Mizuno et al. 2001; Blitz et al. 2007), there is no evidence for a
truncation of the mass spectrum at large masses.  Indeed, the opposite
seems true: the maximum mass clouds are significantly more massive
than predicted by the power-law distribution.  All of the
``overmassive'' clouds are found associated with the star-forming bar
in the LMC and this suggesting a difference in the structure of the
ISM for this region.

Many turbulent models of the ISM predict a log-normal distribution
of object masses (e.g. Ostriker et al. 2001), so we also fit the
cumulative value function of a log-normal distribution to the each
of the catalog distributions:
\begin{equation}
 N(M'>M) = \frac{N_0}{2} \left[ 1+ \mbox{erf} \left(
\frac{\ln M-\ln M_0}{\sigma_{\ln M}}\right)\right].
\end{equation}
The results of the fits are shown in Figure \ref{msp-lognorm}.  The
residuals of the log-normal distribution show significant excursions
outside the counting uncertainties, particularly when compared to the
residuals of the power law fits indicating that a power-law model of
the mass distribution is superior to that of a log-normal distribution
for these data.

Wada, Spaans, and Kim (2000) argued that an index of power law fit can
be steeper if there is star formation ongoing, so the dissipation of
clouds occurs.  In the presence of star formation (their SF2
simulation), they argued that the mass spectrum shows roughly $dN/dM
\propto M^{-2.3}$ for a brightness temperature threshold of 50 K which
is steeper than the comparable index for the 21-cm masses where
$T_{threshold}=64~K$: $\gamma = -1.68\pm 0.04$. Wada et al.~(2000)
also report that $dN/dM \propto M^{-2.7}$ for a brightness temperature
threshold of 30 K and the derived index for $T_{threshold}=32$~K is
$\gamma=-1.65\pm 0.04$.  The resulting catalogs in the LMC have, in
general, a shallower mass distribution than the simulations and there
is not a significant change in the distribution between the 64 and
32~K thresholds.  However, when the threshold is decreased to 16~K,
probing the lowest mass clouds, the index $\gamma \approx -2.0$. For
an index value $\gamma=2$ there is equal mass in each logarithmic bin
of the mass distribution implying the mass structure of the ISM for
low surface brightness clouds have no preferred mass scale.

% ER REVISED
\subsection{The Dynamical State of \ion{H}{1} Clouds}
\label{virparam}
In Figure \ref{vpfig} we compare the virial parameter for \ion{H}{1}
clouds in the the LMC to that of the GMCs.  The virial parameter is
defined so that the dynamical and luminous mass estimates will be
equal.  The virial parameter is plotted as a function of the total
luminous mass.  As seen in Figure \ref{vircomp}, there is no trend in
virial parameter as a function of cloud mass -- all \ion{H}{1} clouds
are in a similar dynamical state.  This is in contrast with the work
of Heyer et al. (2001) who find that, for molecular clouds in the
outer galaxy, high mass clouds show a roughly constant virial
parameters, but low mass clouds follow the locus of pressure
equilibrium with their surroundings.  That the \ion{H}{1} clouds show
no significant change in virial parameter as a function of mass stands
in contrast with the low-mass molecular ``chaff'' seen in the outer
galaxy.  This empirical result should be readily predicted by
simulations of the galactic-scale ISM.

The virial parameter also appears to be constant for the GMCs in the
LMC, except their virial parameter is significantly lower $\langle
\alpha_{GMC}\rangle = 1.8$.  The large margin between the virial
parameters for the two populations of clouds indicates that even the
brightest, most massive components of the LMC highlighted in the
$T_{thresh}=64~$K catalog are not identifying atomic counterparts to
GMCs but rather larger turbulent structures in the ISM.

% ER REVISED
\subsection{The Mass-Radius Relationship}
Since the \ion{H}{1} clouds have roughly constant virial parameters
for a given threshold and also follow a size-line width relationship,
the clouds should also follow a mass-radius relationship.  Such a
relationship is plotted in Figure \ref{mrfig}.  Linear fits to
$\log(M)$ as a function of $\log(R)$ give $M\propto R^{2.05\pm 0.10}$
for all the catalog with the constant being established by the adopted
threshold temperatures. The mean surface densities of the clouds
$\langle \Sigma \rangle = \langle M/\pi R^2\rangle$ are 4.7, 2.5 and
1.3 $M_{\odot}~\mbox{pc}^{-2}$ for $T_{thresh}=64, 32$, and 16~K
respectively, scaling inversely with threshold implying that the
variation is entirely due to the changing threshold values.  Such low
column densities stand in contrast with the column densities of
molecular clouds in the LMC where $\langle \Sigma_{GMC}\rangle =
50~M_{\odot}\mbox{ pc}^{-2}$ \citep{psp5}.

%REVISE
The cumulative size distribution for the identified \ion{H}{1} clouds
is also shown in Figure \ref{sizedist}.  We fit power-law
distributions to the size distribution. For a power law of the form
$dN/dR \propto R^{\beta}$, we find indices of $\beta = -2.2, -2.3,
-3.0$ for the temperature threshold, 64, 32, 16 K, respectively. It is
interesting to note that $\beta \approx -3.0$ at the lowest
temperature threshold is the steepest and similar to that found for
the size spectrum of molecular clouds observed in CO $J=2\rightarrow1$
and CO $J=1\rightarrow0$ (Heithausen et al. 1998; Kramer et al. 1998).

%REVISE
For comparison, cumulative distribution of mean column density
distribution of \ion{H}{1} gas in the LMC along the line of sight are
given in Figure \ref{cdhist} for the individual catalogs and for all
positions in the map in \ref{nhi_pdf}. Neither the cumulative
distribution of \ion{H}{1} surface density nor the probability
distribution function of observed column density along the line of
sight yield a power law fit.  Instead, the column density histogram
appears to have a log-normal distribution instead and we report the
parameters of the log-normal fits in Table \ref{sizedist}.

The column density distributions of small CNM structures studied with
Arecibo (Heiles and Troland 2005) and extracted from numerical
simulation (Hennebelle et al.  2006) follow a power law. Here we also
would like to emphasize that the column density spectrum in the
present study is not well-reproduced by a power law but rather by a
log-normal distribution (in contrast with the mass and radius
distributions which are power laws).  Two features distinguish our
measurement from that of Heiles \& Troland (2005): the structures in
our catalog are significantly larger than those studied in their work
and their observations are based off 21-cm absorption line studies
yielding column densities without any assumptions.

\subsection{Cloud Aspect Ratios}

It is important to note that the present observations also show that
the clumped state of \ion{H}{1} gas appears embedded in the more
diffuse neutral gas clouds. Most clouds are not round, but have an
elongated or filament-like shape. The histogram of measured inverse
aspect ratio of each \ion{H}{1} cloud from the cloud finding algorithm
(Figure \ref{aspect}) indicates that the peak distribution of aspect
ratio arises between 0.3 and 0.4. The mean inverse aspect ratio is about 0.4
and ranges between 0.38 and 0.42. We note that the mean inverse aspect ratio
of the clouds identified with the highest temperature threshold is
slightly larger than that of the clouds detected with lower
temperature threshold.  The presence of smaller and denser \ion{H}{1}
clump-like objects is likely to be comparable with the cold neutral
medium (CNM), one of the four phases of the interstellar medium
(Heiles 2001). As it is now well established, the atomic interstellar
medium is thermally bistable as predicted by detailed computations of
thermal balance (Field, Goldsmith, and Habing 1969; McKee and Ostriker
1977; Wolfire et al. 1995, 2003). \ion{H}{1} gas can be in two
different thermodynamical states. A cold neutral medium (CNM) is
embedded in a warm neutral medium (WNM) and can coexist in thermal
pressure equilibrium (Kulkarni and Heiles 1988).  Recently much
attention has been paid to the smaller scales of \ion{H}{1} emission
and its dynamical properties as a result of extensive surveys using
interferometers and single-dish telescope (Heiles and Troland 2003).
In order to understand the properties of observed \ion{H}{1}, numerous
numerical simulations have been performed (Dib and Burkert 2005;
Gazol, V\'azquez-Semadeni, and Kim 2005; de Avillez and Breitschwerdt
2005). Attempting to resolve the warm and cold components in the
atomic gas, high-resolution numerical simulations were achieved by
various teams (Burkert and Lin 2000; Hennebelle and P\'erault 2000;
Koyama and Inutsuka 2002; Kritsuk and Norman 2002; Piontek and
Ostriker 2005; Audit and Hennebelle 2005; Heitsch et al. 2005;
V\'azquez-Semadeni et al. 2006) and have demonstrated that degree of
turbulence in \ion{H}{1} gas is crucial to understand phase
transformation between the warm diffuse phase and the cold dense phase
induced by dynamical condensation.

\section{Conclusion}

The \ion{H}{1} clouds and clumps have been identified and cataloged
with a brightness temperature threshold ($T_b$) from a 21$-$cm neutral
hydrogen gas survey of the LMC, which was created using the combined
surveys (Kim et al. 2003) of the Australia Telescope Compact Array
(ATCA) (Kim et al. 1998) and the Parkes 64-m single-dish telescope
(Staveley -Smith et al. 2003).  Using our full data set, observations
of \ion{H}{1} establish that the Larson size-line width relation is
obeyed with exponent similar to the Kolmogorov value (Padoan and
Nordlund 2002).  The virial parameters of the clouds are all
significantly larger than unity implying that turbulent energy
dominates the self-gravity of the structures on these size scales.
The \ion{H}{1} clouds distinct from the GMCs in the LMC which have
significantly smaller virial parameters.  The mass and size
distributions of the clouds are both well-represented by power laws
with $dN/dM \propto M^{\alpha}$ with $\alpha = -1.6\to -2.0$ and
$dN/dR \propto R^{\beta}$ with $\beta = -2.5 \to -3.0$.  We find that
the clouds have a roughly constant average column density as a
function of cloud mass, but that column density is distributed around
the mean following a log-normal distribution.  Given the simple nature
of the observations made, the derived distributions should be easily
compared to simulations of the galactic ISM.

\acknowledgements
We thank A. Goodman and E. Keto for interesting discussion. We thank
the anonymous referee for his/her invaluable comments which have
improved the manuscript significantly. SK was supported in part by the
Korea Science and Engineering Foundation (KOSEF), under a cooperative
agreement with the Astrophysical Research Center of the Structure and
Evolution of the Cosmos (ARCSEC). ER's work is supported by an NSF
Astronomy and Astrophysics Postdoctoral Fellowship (AST-0502605).

\providecommand{\bysame}{\leavevmode\hbox
to3em{\hrulefill}\thinspace}
\bibliographystyle{apj}

\begin{deluxetable}{cccccccc}
\tabletypesize{\footnotesize}
\tablecaption{Catalog of \ion{H}{1} cloud for the brightness temperature
threshold of 16 K}
\tablewidth{0pt}
\tablenum{1}
\tablehead{\colhead{ID} & \colhead{RA} & \colhead{DEC} & \colhead{SIZE} & \colhead{Vel} & \colhead{$\sigma$} & \colhead{$M_{cloud} $}
& \colhead{REGION} \\
\colhead{} & \colhead{(J2000)} & \colhead{(J2000)} &
\colhead{(pc)} & \colhead{($\rm{ kms^{-1}} $)} & \colhead{($\rm{ kms^{-1}} $)} &
\colhead{($M_\odot$)}
& \colhead{}  }
\startdata
  1 & 4:30:39.60 & -69:05:24.00 &   51.180 &  251.71 &  1.456 &         9601.75 &                 \\
  2 & 4:33:11.04 & -69:17:60.00 &   21.948 &  247.22 &  1.996 &         2178.55 &                 \\
  3 & 4:34:10.32 & -69:22:19.20 &   19.437 &  253.31 &  1.242 &         1452.37 &                 \\
  4 & 4:34:30.24 & -67:16:51.60 &   14.845 &  265.00 &  1.594 &         1048.93 &                 \\
  5 & 4:35:16.56 & -68:59:38.40 &   17.142 &  249.02 &  1.465 &         1129.62 &                 \\
  6 & 4:35:23.28 & -67:13:40.80 &   20.283 &  258.75 &  1.926 &         2178.55 &                 \\
  7 & 4:35:36.48 & -70:52:44.40 &   15.969 &  236.28 &  0.449 &          645.50 &                 \\
  8 & 4:35:43.20 & -69:09:25.20 &   21.971 &  244.99 &  1.433 &         1452.37 &                 \\
\enddata
\end{deluxetable}

\begin{deluxetable}{cccccccc}
\tabletypesize{\footnotesize}
\tablecaption{Catalog of \ion{H}{1} cloud for the brightness temperature
threshold of 32 K}
\tablenum{2}
\tablewidth{0pt}
\tablehead{\colhead{ID} & \colhead{RA} & \colhead{DEC} & \colhead{SIZE} & \colhead{Vel} & \colhead{$\sigma$} & \colhead{$M_{cloud} $} & \colhead{REGION}\\
\colhead{} & \colhead{(J2000)} & \colhead{(J2000)} &
\colhead{(pc)} & \colhead{($\rm{ kms^{-1}} $)} & \colhead{($\rm{ kms^{-1}} $)} &
\colhead{($M_\odot$)} & \colhead{} }
\startdata
  1 & 4:37:49.44 & -66:18:46.80 &   33.682 &  272.59 &  1.959 &         8230.07 &                 \\
  2 & 4:38:36.00 & -67:05:60.00 &   41.914 &  265.07 &  4.509 &        13232.67 &                 \\
  3 & 4:38:56.16 & -66:58:19.20 &   43.366 &  261.68 &  1.612 &         7988.01 &                 \\
  4 & 4:39:02.64 & -69:27:21.60 &   37.388 &  245.36 &  1.578 &         5406.03 &                 \\
  5 & 4:39:36.00 & -67:12:54.00 &   37.834 &  267.78 &  1.710 &         9601.75 &                 \\
  6 & 4:39:49.44 & -67:35:31.20 &   17.325 &  267.51 &  0.871 &         1129.62 &                 \\
  7 & 4:40:02.88 & -67:45:39.60 &   55.148 &  266.07 &  1.490 &        15249.84 &    Northern Arm \\
  8 & 4:40:09.60 & -67:19:44.40 &   35.957 &  264.77 &  1.173 &         8714.20 &    Northern Arm \\
\enddata
\end{deluxetable}

\begin{deluxetable}{cccccccc}
\tabletypesize{\footnotesize} \tablecaption{Catalog of \ion{H}{1} cloud
for the brightness temperature threshold of 64 K} \tablenum{3}
\tablewidth{0pt} \tablehead{\colhead{ID} & \colhead{RA} &
\colhead{DEC} & \colhead{SIZE} & \colhead{Vel} &
\colhead{$\sigma$} & \colhead{$M_{cloud} $} & \colhead{REGION}\\
\colhead{} & \colhead{(J2000)} & \colhead{(J2000)} &
\colhead{(pc)} & \colhead{($\rm{ kms^{-1}} $)} & \colhead{($\rm{kms^{-1}} $)} & \colhead{($M_\odot$)} & \colhead{} }
\startdata
  1 & 4:42:57.12 & -68:41:56.40 &   18.945 &  256.27 &  1.382 &         4437.78 &    Northern Arm \\
  2 & 4:43:37.44 & -68:49:22.80 &   20.994 &  258.43 &  1.067 &         5163.97 &    Northern Arm \\
  3 & 4:43:44.16 & -67:15:10.80 &   44.371 &  272.13 &  1.500 &        16056.71 &    Northern Arm \\
  4 & 4:44:51.36 & -69:01:01.20 &   23.584 &  257.13 &  2.054 &         7503.89 &    Northern Arm \\
  5 & 4:46:18.96 & -68:19:33.60 &   19.917 &  273.55 &  1.040 &         4195.72 &    Northern Arm \\
  6 & 4:46:59.52 & -68:49:01.20 &   52.636 &  263.16 &  2.640 &        26303.96 &    Northern Arm \\
  7 & 4:47:33.36 & -69:15:46.80 &   15.536 &  251.83 &  1.068 &         3308.17 &    Northern Arm \\
  8 & 4:48:13.92 & -68:23:13.20 &   69.132 &  264.65 &  1.521 &        33807.85 &    Northern Arm \\
\enddata
\end{deluxetable}

\begin{deluxetable}{cccc}
\tablecaption{Derived Parameters of the Mass
Distributions\label{mspec-index}} \tablewidth{0pt} \tablenum{4}
\tablehead{ \colhead{Parameter} & \colhead{$T_{thresh}=64$~K} &
\colhead{$T_{thresh}=32$~K} & \colhead{$T_{thresh}=16$~K}}
\startdata \cutinhead{Power-Law Distributions} Completeness Limit
($M_{\odot}$) & $10^4$ &  $5\times 10^3$ &
$2.5\times 10^3$\\
$N_0$ &  $22 \pm 15 $& $34 \pm 16$ & $5 \pm 2$ \\
$\gamma$ &  $-1.68 \pm 0.04$ & $-1.65 \pm 0.04$  & $-1.99\pm 0.04$ \\
\cutinhead{Log-normal distribution.}
$N_0$ & $800\pm 200$ & $1400 \pm 150$ & $810\pm 30$ \\
$\sigma_{\ln M}$ & $4.2\pm 0.3$ &
$4.1\pm 0.1$ & $2.34 \pm 0.04$ \\
$\ln(M_0/M_{\odot})$ & $5.9\pm 0.7$  & $5.7\pm 0.3$ & $6.8\pm 0.1$\\
\enddata
\end{deluxetable}

\begin{deluxetable}{cccc}
\tablecaption{Derived Parameters of the Column Density
Distributions\label{cdparams}} \tablewidth{0pt} \tablenum{5}
\tablehead{ \colhead{Parameter} & \colhead{$T_{thresh}=64$~K} &
\colhead{$T_{thresh}=32$~K} & \colhead{$T_{thresh}=16$~K}}
\startdata
$N_0$ & $200\pm 10$ & $430 \pm 20$ & $480\pm 20$ \\
$\sigma_{\ln N(\mathrm{H})}$ & $0.64\pm 0.05$ &
$0.78\pm 0.04$ & $0.78 \pm 0.04$ \\
$(N_{\mathrm{H},0}/10^{20}\mbox{ cm}^{-2})$ & $6.0\pm 0.4$  & $3.1\pm 0.3$ & $1.6\pm 0.2$\\
\enddata
\end{deluxetable}

\begin{figure}
\centerline{\includegraphics[width=1.0\textwidth]{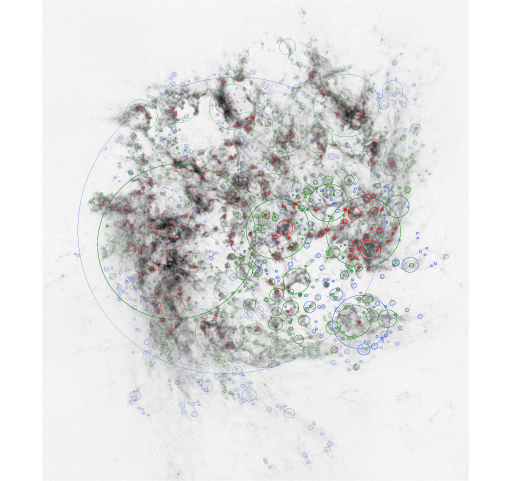}}
%\plotone{fig1.eps}
\caption[distribution of HI clouds]{The size
distribution of \ion{H}{1} clouds in the LMC. Size of an ellipse
presents size of \ion{H}{1} cloud. Blue, green, and red colors
represent brightness temperature threshold at 16 K, 32 K, and 64 K
respectively.}
\end{figure}

\begin{figure}
\centerline{\includegraphics[width=1.0\textwidth]{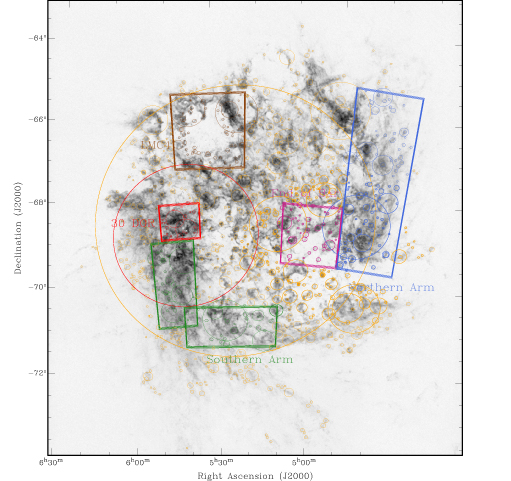}}
%\plotone{fig2.eps}
\caption[distribution of HI clouds]{2D spatial
distribution and size distribution of clouds selected for each
region.}
\end{figure}

\begin{figure} \epsscale{1.0}
\centerline{\includegraphics[width=1.0\textwidth]{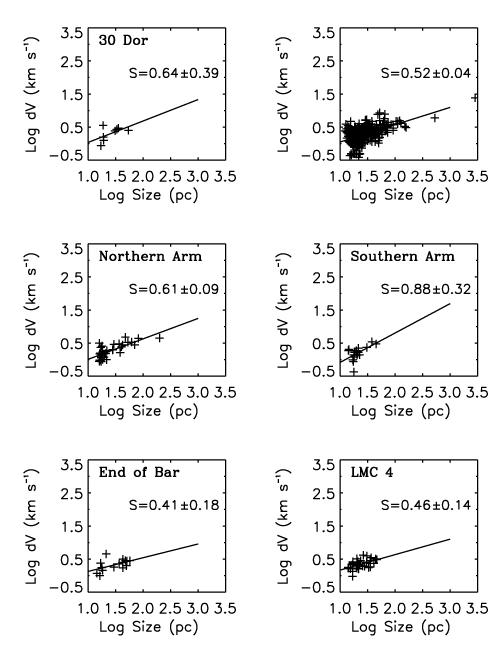}}
\caption[size vs line width diagram] {A plot of the size versus the
velocity dispersion of \ion{H}{1} cloud for each region ($T_b = 16 K
$). 30 Dor, Northern \& Southern Spiral Arm, End of Bar, and the LMC
4 indicate the dynamically hot regions and the relatively quiet
region is presented in the upper right panel.}
\end{figure}

\begin{figure}
\centerline{\includegraphics[width=1.0\textwidth]{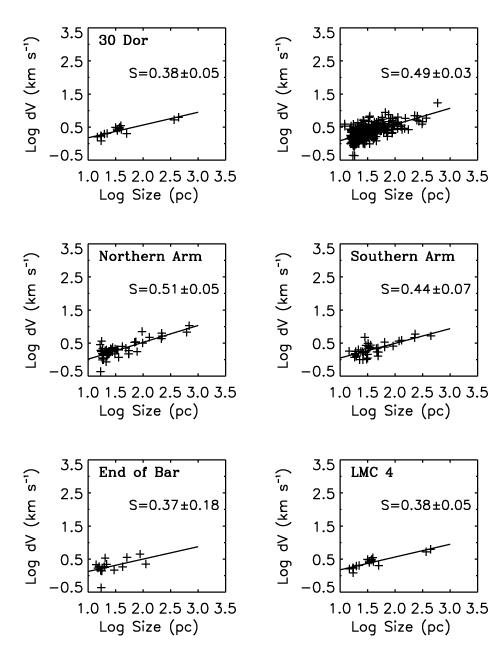}}
\caption[size vs line width diagram] {A plot of the size versus the
velocity dispersion of \ion{H}{1} cloud for each region ($T_b = 32 K
$). 30 Dor, Northern \& Southern Spiral Arm, End of Bar, and the LMC
4 indicate the dynamically hot regions and the relatively quiet
region is presented in the upper right panel.}
\end{figure}

\begin{figure}
\centerline{\includegraphics[width=1.0\textwidth]{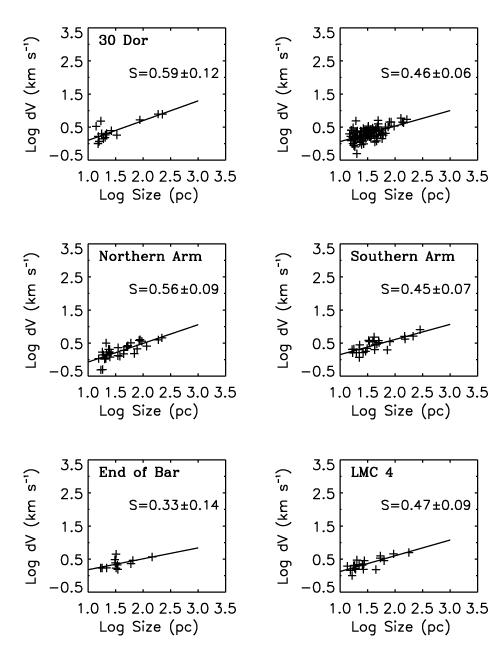}}
\caption[size vs line width diagram] {A plot of the size versus the
velocity dispersion of \ion{H}{1} cloud for each region ($T_b = 64 K
$). 30 Dor, Northern \& Southern Spiral Arm, End of Bar, and the LMC
4 indicate the dynamically hot regions and the relatively quiet
region is presented in the upper right panel.}
\end{figure}

\begin{figure}
\centerline{\includegraphics[width=1.0\textwidth]{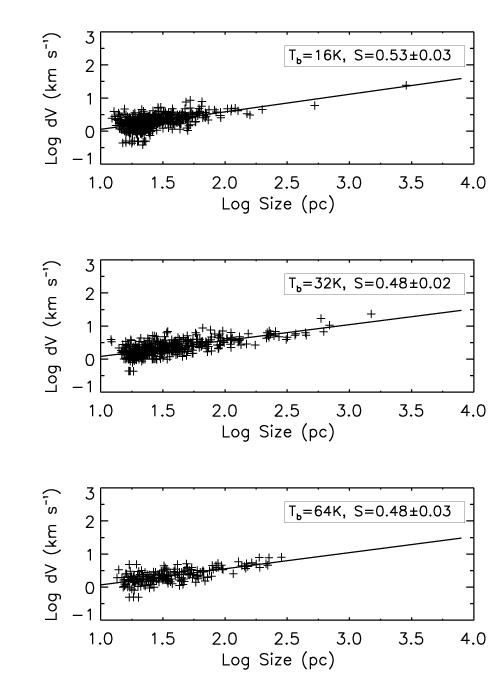}}
\caption[Size$-$Line Width Relation of the HI Gas]{Size$-$Line Width
Relation of the \ion{H}{1} clouds in the entire region of the LMC
for a given brightness temperature threshold.\label{rdv}}
\end{figure}

\begin{figure}
\centerline{\includegraphics[width=1.0\textwidth]{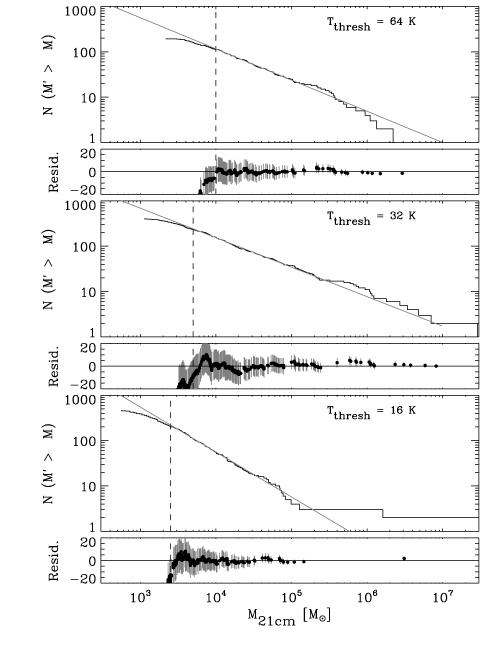}}
\vspace*{-1.0cm} \caption{\label{mspec} Cumulative mass distribution
based on 21-cm luminosity for $T_{threshold}=64,32,$ and $16$ K in
the top, middle and bottom large panels respectively.  For each
distribution, we fit a power law above masses of $10^4~M_{\odot}$
which is shown as a gray line. Parameters of the fit are given in
Table 4.  A vertical dashed line indicates the lower limit of clouds
included in the fit. Small panels below each of the large panels
represent the residuals of the fit in each case.}
\end{figure}

% ER REVISED
\begin{figure}
\centerline{\includegraphics[width=1.0\textwidth]{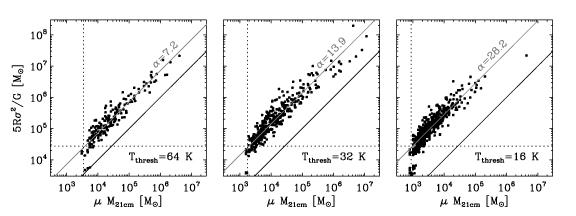}}
%\includegraphics[width=1.0\textwidth]{plotLMC.eps}
%\plotone{vircomp.ps}
\caption{\label{vircomp} Comparison of luminous and dynamical mass
estimates for clouds in each of the catalogs. The median virial
parameter for each catalog is indicated with the gray diagonal line.
The completeness limits for both quantities are shown as dotted
lines. }
\end{figure}

\begin{figure}
\centerline{\includegraphics[width=1.0\textwidth]{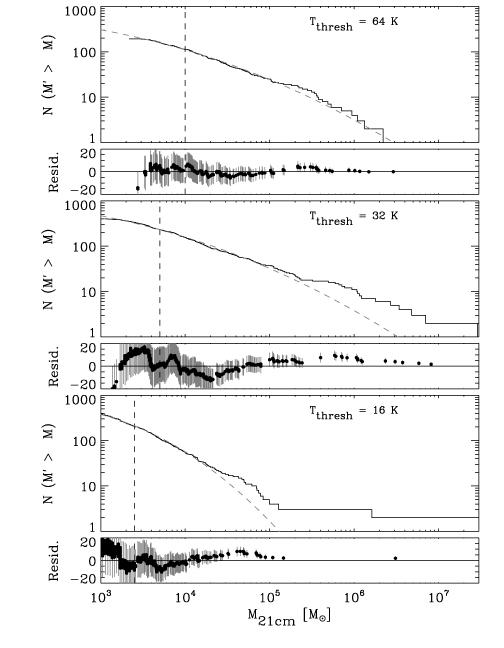}}
\vspace*{-1.0cm}\caption{\label{msp-lognorm} Cumulative mass
 distribution based on 21-cm luminosity for $T_{threshold}=64,32,$ and
 $16$ K in the top, middle and bottom large panels respectively.  For
 each distribution, we fit a cumulative value function for a
 log-normal distribution above masses of $10^4~M_{\odot}$ which is
 shown as a gray line.  Parameters of the fit are given in Table 4.
 A vertical dashed line indicates the lower limit of clouds included
 in the fit.  Small panels below each of the large panels represent
 the residuals of the fit in each case. }
\end{figure}

% ER REVISED
\begin{figure}
%vp.ps
\centerline{\includegraphics[width=1.0\textwidth]{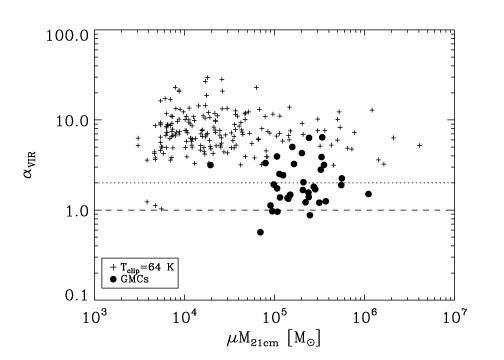}}
\caption{\label{vpfig} Virial parameter as a function of luminous
  mass for \ion{H}{1} clouds and GMCs in the LMC.  Neither type of
  clouds shows an obvious trend in the virial parameter with mass, but
  the mean virial parameters of the two systems are quite different. }
\end{figure}

%ER REVISED
\begin{figure}
%\plotone{mr.ps}
\centerline{\includegraphics[width=1.0\textwidth]{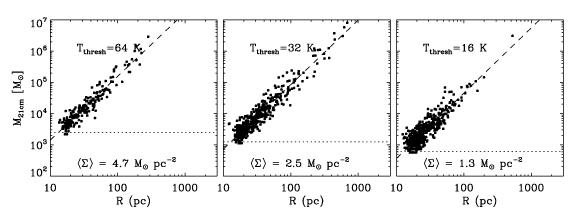}}
\caption{\label{mrfig} Radius-mass relationship for \ion{H}{1}
clouds
 in the LMC.  The clouds follow the $M\propto R^{2}$ scaling expected
 from Larson's laws based on a constant virial parameter and a
 size-line width relationship.  Hence, the mean surface densities for
 all clouds are roughly constant and given in the panels of the Figure.}
\end{figure}

\begin{figure}
%\plottwo{cdhist.ps}{nhi_pdf.ps}
\centerline{\includegraphics[width=1.0\textwidth]{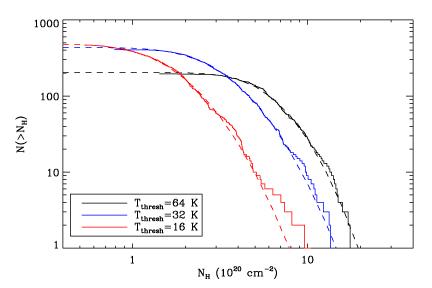}}
\caption{\label{cdhist} Cumulative distribution of \ion{H}{1} column
  density.  The observed distribution of column densities is shown as
  a solid line and a log-normal fit to the distribution is shown as a
  dashed line (see Table \ref{cdparams} for parameters).  A power-law
  distribution would not reproduce the observed distribution.}
\end{figure}

\begin{figure}
\centerline{\includegraphics[width=1.0\textwidth]{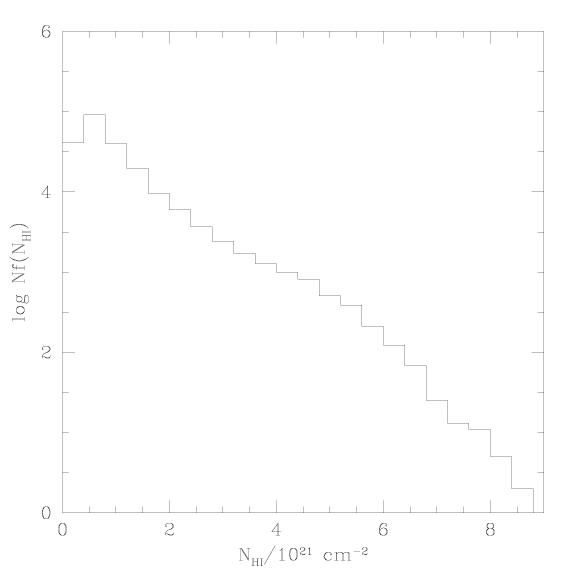}}
\caption{\label{nhi_pdf}Observed column density distribution of
\ion{H}{1} along the line of sight is given.}
\end{figure}

\begin{figure}
\centerline{\includegraphics[width=1.0\textwidth]{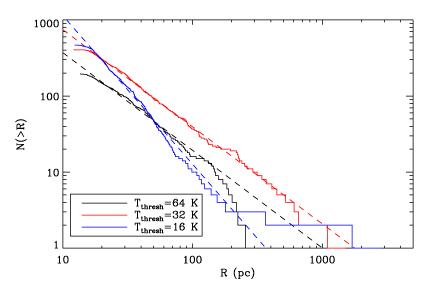}}
\caption{\label{sizedist} Size distribution of \ion{H}{1} clouds. A
power law form, $dN/dR \propto R^{\beta}$, is fit to the size
spectrum for each catalog of clouds reproducing the observed shape
of the distribution over the entire data range.}
\end{figure}

\begin{figure}
\centerline{\includegraphics[width=1.0\textwidth]{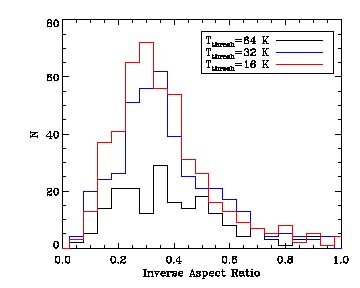}}
\caption{Histogram of inverse aspect ratio (minor/major) of the
clouds.  The peak distribution of inverse aspect ratio increases to
0.4 for the lower brightness temperature threshold.\label{aspect}}
\end{figure}
\end{document}